\documentclass[aps,prd,onecolumn,groupedaddress,showpacs,nofootinbib,amssymb]{revtex4}
\usepackage[dvips]{graphicx}
\usepackage{amssymb}
\usepackage{amsmath}
\usepackage{graphicx,,color}
\usepackage{amsfonts}
\usepackage{bm}
\usepackage{cancel}
\usepackage{comment}

\newcommand\be{\begin{equation}}
\newcommand\ee{\end{equation}}
\newcommand\nn{\nonumber \\}
\newcommand\e{\mathrm{e}}

\allowdisplaybreaks[4]

\begin{document}

\tolerance=5000

\title{Ghost-free $F\left( R,\mathcal{G} \right)$ Gravity}
\author{S.~Nojiri,$^{1,2}$}
\email{nojiri@gravity.phys.nagoya-u.ac.jp}
\author{S.~D.~Odintsov,$^{3,4,5}$}
\email{odintsov@ieec.uab.es}
\author{V.K.~Oikonomou,$^{6,7}$}
\email{v.k.oikonomou1979@gmail.com}
\author{Arkady A. Popov,$^{8}$}
\email{arkady_popov@mail.ru}
\affiliation{$^{1)}$ Department of Physics, Nagoya University, Nagoya 464-8602, Japan \\
$^{2)}$ Kobayashi-Maskawa Institute for the Origin of Particles
and the Universe, Nagoya University, Nagoya 464-8602, Japan \\
$^{3)}$ ICREA, Passeig Luis Companys, 23, 08010 Barcelona, Spain\\
$^{4)}$ Institute of Space Sciences (IEEC-CSIC) C. Can Magrans
s/n, 08193 Barcelona, Spain\\
$^{5)}$ Institute of Physics, Kazan Federal University,  Kazan
420008, Russia\\
$^{6)}$ Department of Physics, Aristotle University of
Thessaloniki, Thessaloniki 54124, Greece\\
$^{7)}$ Laboratory for Theoretical Cosmology, Tomsk State
University of Control Systems and Radioelectronics, 634050 Tomsk, Russia (TUSUR)\\
$^{8)}$ N. I. Lobachevsky Institute of Mathematics and Mechanics,
Kazan Federal University,\\
420008, Kremlevskaya street 18, Kazan, Russia\\
}

\tolerance=5000

\begin{abstract}
In this work we shall address the ghost issue of
$F\left( R,\mathcal{G} \right)$ gravity, which is known to be plagued with
ghost degrees of freedom. These ghosts occur due to the presence
of higher than two derivatives in the field equations, and can
arise even when considering cosmological perturbations, where
superluminal modes may arise in the theory.
If we consider the quantum theory, the ghosts generate the negative norm states,
which give the negative probabilities, and therefore the ghosts are physically inconsistent.
Motivated by the
importance of $F\left( R,\mathcal{G} \right)$ gravity for providing viable
inflationary and dark energy phenomenologies, in this work we
shall provide a technique that can render $F\left( R,\mathcal{G} \right)$
gravity theories free from ghost degrees of freedom. This will be
done by introducing two auxiliary scalar fields, and by employing
the Lagrange multiplier technique, the theory is ghost free in the
Einstein frame. Also the framework can be viewed as a
reconstruction technique and can be used as a method in order to
realize several cosmological evolutions of interest. We
demonstrate how we can realize several cosmologically interesting phenomenologies
by using the reconstruction technique.
\end{abstract}

\pacs{04.50.Kd, 95.36.+x, 98.80.-k, 98.80.Cq,11.25.-w}

\maketitle

\section{Introduction \label{SecI}}

Modified gravity in its various forms
\cite{Nojiri:2017ncd,Nojiri:2010wj,Nojiri:2006ri,Capozziello:2011et,Faraoni:2010pgm,
delaCruz-Dombriz:2012bni,Olmo:2011uz} plays a prominent role
towards the complete understanding of how the Universe evolves,
due to the fact that Einstein-Hilbert gravity seems to be unable
to describe the late-time acceleration era
\cite{Bamba:2012cp,Peebles:2002gy,Li:2011sd,Bamba:2010wb,Frieman:2008sn,
Boehmer:2008av,Sahni:2006pa,Nojiri:2006gh,Elizalde:2004mq,Makarenko:2018blx,
Capozziello:2003gx,Kamenshchik:2001cp,Carroll:1998zi,Capozziello:2002rd,Capozziello:2005ra}.
Apart from the dark energy description, modified gravity can
provide a viable and perhaps necessary description for the early
time acceleration dubbed inflation
\cite{Guth:1980zm,Linde:1993cn,Linde:1983gd}. Single scalar field
descriptions of inflation for the moment are quite popular, but
these models provide an inflationary era with specific
characteristics and there is not much freedom in model building in
these theories. Specifically, single scalar field models lead to a
Gaussian power spectrum which can be compatible with the latest
Planck data \cite{Planck:2018jri}, however the tensor spectral
index is red-tilted and obeys the consistency relation, a fact
that restricts too much the tensor spectrum. If in the upcoming
observational data of the LISA mission
\cite{Baker:2019nia,Smith:2019wny} in about fifteen years from
now, primordial gravitational waves signal will be found, single
scalar field inflation will be instantly unable to provide a
description for the early time era evolution of our Universe. This
is due to the fact that currently the single scalar field
prediction for the power spectrum of the primordial gravitational
waves is way lower compared to the LISA sensitivity curve. Thus if
a signal of primordial tensor spectrum is discovered, single
scalar field cannot be the one generating this spectrum, unless
the scalar field is a tachyon, which is quite problematic
scenario. These issues render modified gravity a quite timely and
appealing candidate for the unified description of inflation and
dark energy simultaneously. The first work toward this direction
was performed in the pioneer work \cite{Nojiri:2003ft}. Two of the
most appealing characteristic modified gravities are $f(R)$ and
Gauss-Bonnet gravity
\cite{Li:2007jm,Nojiri:2005jg,Nojiri:2005am,Cognola:2006eg,Elizalde:2010jx,Izumi:2014loa,
Oikonomou:2016rrv,Oikonomou:2015qha,Escofet:2015gpa,Makarenko:2017vuk,Bamba:2014mya,Makarenko:2016jsy}
and higher order extensions of the above two
\cite{Clifton:2006kc,Bogdanos:2009tn,Capozziello:2004us,Barrow:1988xh,Elizalde:2010jx,
Bamba:2009uf,DeLaurentis:2015fea,Benetti:2018zhv,delaCruzDombriz:2011wn,deMartino:2020yhq,SantosDaCosta:2018bbw}.
In this work we shall consider $F\left( R,\mathcal{G} \right)$
gravity theories
\cite{Elizalde:2010jx,Bamba:2009uf,DeLaurentis:2015fea,Benetti:2018zhv,delaCruzDombriz:2011wn,
deMartino:2020yhq,SantosDaCosta:2018bbw}, and we shall consider
the problem of having ghost degrees of freedom in these theories.
It is known that $F\left( R,\mathcal{G} \right)$ gravity theories
are plagued with ghost degrees of freedom which can occur at many
levels in the theory, even as perturbative propagating modes when
cosmological perturbations are considered. Furthermore if we
consider the quantum theory, the ghosts generate the negative norm
states, which give the negative probabilities, and therefore the
ghosts are physically inconsistent. We aim to demonstrate how
these ghost degrees of freedom are generated and we will provide a
theoretical framework for $F\left( R,\mathcal{G} \right)$ which
can yield a ghost-free framework. We also use the provided
theoretical framework as a reconstruction technique in order to
realize several cosmological evolutions of interest.

The paper is organized as follows: In section \ref{SecII}, we demonstrate
how ghost degrees of freedom can arise in $F\left( R,\mathcal{G} \right)$ gravity.
In section \ref{SecIII}, we provide a ghost-free theoretical
framework for $F\left( R,\mathcal{G} \right)$ gravity and in section \ref{SecIV}, we
describe in detail the cosmological framework of the ghost-free
$F\left( R,\mathcal{G} \right)$ gravity, providing also the reconstruction
technique for it.
In section \ref{SecV}, we use the reconstruction technique
for realizing several cosmological evolutions of interest, and
finally the conclusions follow in the end of the article.

Throughout this paper we use the conventions for the curvature
tensor $R^\lambda_{\ \mu\rho\nu}= -\Gamma^\lambda_{\mu\rho,\nu} +
\Gamma^\lambda_{\mu\nu,\rho} -
\Gamma^\eta_{\mu\rho}\Gamma^\lambda_{\nu\eta} +
\Gamma^\eta_{\mu\nu}\Gamma^\lambda_{\rho\eta}$ and for the Ricci
tensor $R_{\mu\nu}= R^\lambda_{\ \mu\lambda\nu}$.

\section{Occurrence of Ghost Degrees of Freedom in $F\left( R, \mathcal{G} \right)$ Gravity \label{SecII}}

In this section, we shall demonstrate that the $F\left( R, \mathcal{G} \right)$
has inherent ghost degrees of freedom.
The $F\left( R, \mathcal{G} \right)$ gravity in vacuum has the
following action,
\begin{equation}
\label{FRGBg1} S = \int d^4 x \sqrt{-g} F\left( R, \mathcal{G} \right)\, ,
\end{equation}
where $F\left( R, \mathcal{G} \right)$ is a function of the scalar
curvature $R$ and $\mathcal{G}$ stands for the Gauss-Bonnet
invariant given by,
\begin{equation}
\label{GB}
\mathcal{G} \equiv R^2 - 4 R_{\mu\nu} R^{\mu\nu} + R_{\mu\nu\rho\sigma} R^{\mu\nu\rho\sigma} \, .
\end{equation}
In general, the model (\ref{FRGBg1}) leads to ghost instabilities
and ghost degrees of freedom, that eventually appear even at the level of cosmological perturbations.
As an explicit example, we investigate the so-called $f(\mathcal{G})$ gravity, which is a
special model of $F\left( R, \mathcal{G} \right)$ gravity, as in
\cite{Nojiri:2018ouv}, with action,
\begin{equation}
\label{GB1b}
S=\int d^4x\sqrt{-g} \left(\frac{1}{2\kappa^2}R +f(\mathcal{G}) + \mathcal{L}_\mathrm{matter}\right)\, .
\end{equation}
Upon variation of the action with respect to the metric, we obtain
the following equation of motion,
\begin{align}
\label{fRGB7}
0 = & \frac{1}{2\kappa^2}\left(- R_{\mu\nu} + \frac{1}{2}g_{\mu\nu} R\right)
+ \frac{1}{2} T_{\mathrm{matter}\, \mu\nu} + \frac{1}{2}g_{\mu\nu} \left( f(\mathcal{G})
 - \mathcal{G} f' \left( \mathcal{G} \right) \right)
+ D_{\mu\nu}^{\ \ \tau\eta} \nabla_\tau \nabla_\eta f' \left( \mathcal{G} \right) \, , \nn
D_{\mu\nu}^{\ \ \tau\eta} \equiv& \left( \delta_\mu^{\ \tau} \delta_\nu^{\ \eta}
+ \delta_\nu^{\ \tau} \delta_\mu^{\ \eta} - 2 g_{\mu\nu} g^{\tau\eta} \right) R
+ \left( - 4 g^{\rho\tau} \delta_\mu^{\ \eta} \delta_\nu^{\ \sigma}
 - 4 g^{\rho\tau} \delta_\nu^{\ \eta} \delta_\mu^{\ \sigma}
+ 4 g_{\mu\nu} g^{\rho\tau} g^{\sigma\eta} \right) R_{\rho\sigma} \nn
& + 4 R_{\mu\nu} g^{\tau\eta} - 2 R_{\rho\mu\sigma \nu} \left( g^{\rho\tau} g^{\sigma\eta}
+ g^{\rho\eta} g^{\sigma\tau} \right) \, .
\end{align}
Let a solution of (\ref{fRGB7}) be $g_{\mu\nu}=g^{(0)}_{\mu\nu}$
and we denote the curvatures and connections given by
$g^{(0)}_{\mu\nu}$ by using the indexes ``$(0)$''. Then in order
to investigate if any ghost could exist, we may consider the
perturbation of (\ref{fRGB7}) around the solution
$g^{(0)}_{\mu\nu}$ as follows $g_{\mu\nu}=g^{(0)}_{\mu\nu}
+ \delta g_{\mu\nu}$. For the variation of $\delta g_{\mu\nu}$, we
may impose the transverse gauge condition
$0 = \nabla^\mu \delta g_{\mu\nu}$, and further if we impose the condition
$\delta g^\mu_{\ \mu}=0$, we find
\begin{equation}
\label{FRGBg10}
\delta\mathcal{G}= - 2 R R^{\mu\nu} \delta g_{\mu\nu}
+ 8 R^{\rho\sigma} R^{\mu\ \nu}_{\ \rho\ \sigma}\delta g_{\mu\nu}
+ 4 R^{\mu\nu} \nabla^2 \delta g_{\mu\nu}
 - 2 R^{\mu\rho\sigma\tau} R^\nu_{\ \rho\sigma\tau} \delta g_{\mu\nu}
 - 4 R^{\rho\mu\sigma\nu} \nabla_\rho \nabla_\sigma \delta g_{\mu\nu} \, ,
\end{equation}
which also contains the second derivative of the metric $g_{\mu\nu}$ with respect the cosmic time coordinate.
Under the perturbation $g_{\mu\nu}=g^{(0)}_{\mu\nu} + \delta g_{\mu\nu}$,
the term $D_{\mu\nu}^{\ \ \tau\eta} \nabla_\tau \nabla_\eta f' \left( \mathcal{G} \right)$ takes the following form,
\begin{equation}
\label{FRGBg11}
D_{\mu\nu}^{\ \ \tau\eta} \nabla_\tau \nabla_\eta f' \left( \mathcal{G} \right)
\to D_{\mu\nu}^{\ \ \tau\eta} \nabla_\tau \nabla_\eta f' \left( \mathcal{G}^{(0)} \right)
+ D_{\mu\nu}^{\ \ \tau\eta} \nabla_\tau \nabla_\eta
\left( f'' \left( \mathcal{G}^{(0)} \right) \delta \mathcal{G} \right) + \cdots \, ,
\end{equation}
which contains the fourth derivative of the metric $g_{\mu\nu}$
with respect to the cosmic time coordinate, and therefore the
perturbed equation (\ref{fRGB7}) will have a ghost mode.
Note that in Eq.~(\ref{FRGBg11}), the ``$\cdots$'' expresses the terms
occurring from the variation of $D_{\mu\nu}^{\ \ \tau\eta} \nabla_\tau \nabla_\eta$.
The propagating mode is a scalar
expressed by the Gauss-Bonnet invariant as it is clear from
Eq.~(\ref{fRGB7}).

\section{Ghost-Free $F\left( R, \mathcal{G} \right)$ gravity \label{SecIII}}

We review on the construction of the ghost-free $F\left( R, \mathcal{G} \right)$
theory of gravity based on \cite{Nojiri:2018ouv}.
By introducing two auxiliary fields $\Phi$ and $\Theta$, the action of
Eq.~(\ref{FRGBg1}) can be rewritten as follows,
\begin{equation}
\label{FRGBg2}
S = \int d^4 x \sqrt{-g} \left\{ \frac{\Phi R}{2\kappa^2} + \Theta \mathcal{G}
 - V\left( \Phi, \Theta \right) \right\}\, ,
\end{equation}
where we have introduced the gravitational coupling $\kappa$ in
order to make $\Phi$ and $\Theta$ dimensionless.
By varying the action (\ref{FRGBg2}) with respect to the scalar fields $\Phi$ and
$\Theta$, we obtain,
\begin{equation}
\label{FRGBg3}
\frac{R}{2\kappa^2} = \frac{\partial V\left( \Phi, \Theta \right)}{\partial \Phi} \, ,
\quad \mathcal{G} = \frac{\partial V\left( \Phi, \Theta \right)}{\partial \Theta} \, ,
\end{equation}
which can be algebraically solved with respect $\Phi$ and
$\Theta$, that is, $\Phi = \Phi \left( R, \mathcal{G} \right)$ and
$\Theta = \Theta \left( R, \mathcal{G} \right)$.
Then by substituting the obtained expressions for
$\Phi = \Phi \left( R, \mathcal{G} \right)$ and
$\Theta = \Theta \left( R, \mathcal{G} \right)$ in Eq.~(\ref{FRGBg2}),
we obtain the action, (\ref{FRGBg1}) with,
\begin{equation}
\label{FRGBg4} F\left( R, \mathcal{G} \right) = \frac{\Phi \left(
R, \mathcal{G} \right) R}{2\kappa^2} + \Theta \left( R,
\mathcal{G} \right) \mathcal{G}
 - V\left( \Phi \left( R, \mathcal{G} \right),
\Theta \left( R, \mathcal{G} \right) \right) \, .
\end{equation}
In order to investigate the properties of the action
(\ref{FRGBg2}), we work in the Einstein frame, so under a
conformal transformation of the form $g_{\mu\nu}\to \e^\phi g_{\mu\nu}$,
the curvatures are transformed as follows~\cite{Maeda:1988ab,Nojiri:2003ft},
\begin{align}
\label{E4}
R_{\zeta\mu\rho\nu} \to& \left\{R_{\zeta\mu\rho\nu}
 - \frac{1}{2}\left(g_{\zeta\rho}\nabla_\nu \nabla_\mu \phi
+ g_{\mu\nu}\nabla_\rho \nabla_\zeta \phi
 - g_{\mu\rho} \nabla_\nu \nabla_\zeta \phi
 - g_{\zeta\nu} \nabla_\rho \nabla_\mu \phi \right) \right. \nn
& + \frac{1}{4}\left(g_{\zeta\rho}\partial_\nu \phi \partial_\mu \phi
+ g_{\mu\nu} \partial_\rho \phi \partial_\zeta \phi
 - g_{\mu\rho} \partial_\nu \phi \partial_\zeta \phi
 - g_{\zeta\nu} \partial_\rho \phi \partial_\mu \phi \right) \nn
& - \frac{1}{4}\left(g_{\zeta\rho} g_{\mu\nu}
 - g_{\zeta\nu} g_{\mu\rho}\right)\partial^\sigma \phi
\partial_\sigma \phi \Bigr\}\, ,\nn
R_{\mu\nu} \to& R_{\mu\nu} - \frac{1}{2}\left( 2\nabla_\mu \nabla_\nu \phi
+ g_{\mu\nu} \Box \phi\right)
+ \frac{1}{2}\partial_\mu \phi \partial_\nu \phi - \frac{1}{2} g_{\mu\nu}\partial^\sigma \phi
\partial_\sigma \phi\, ,\nn
R \to& \left(R - 3 \Box \phi - \frac{3}{2}\partial^\sigma \phi \partial_\sigma \phi
\right)\e^{-\phi}\, .
\end{align}
Therefore the Gauss-Bonnet invariant $\mathcal{G}$ is transformed in the following way,
\begin{equation}
\label{fG6}
\mathcal{G} \to \e^{-2\phi} \left[ \mathcal{G} + \nabla_\mu \left\{
4 \left( R^{\mu\nu} - \frac{1}{2} g^{\mu\nu} R \right) \partial_\nu \phi
+ 2 \left( \partial^\mu \phi \Box \phi
 - \left( \nabla_\nu \nabla^\mu \phi\right) \partial^\nu \phi \right)
+ \partial_\nu \phi \partial^\nu \phi \partial^\mu \phi \right\} \right] \, .
\end{equation}
Then by writing $\Phi=\e^{-\phi}$, the action of
Eq.~(\ref{FRGBg2}) can be rewritten by taking into account the
conformal transformation $g_{\mu\nu}\to \e^\phi g_{\mu\nu}$ as follows,
\begin{align}
\label{FRGBg4b}
S =& \int d^4 x \sqrt{-g} \left\{ \frac{1}{2\kappa^2} \left( R
 - \frac{3}{2}\partial^\sigma \phi \partial_\sigma \phi \right) \right. \nn
& \left. + \Theta \mathcal{G} - \partial_\mu \Theta \left\{
4 \left( R^{\mu\nu} - \frac{1}{2} g^{\mu\nu} R \right) \partial_\nu \phi
+ 2 \left( \partial^\mu \phi \Box \phi
 - \left( \nabla_\nu \nabla^\mu \phi\right) \partial^\nu \phi \right)
+ \partial_\nu \phi \partial^\nu \phi \partial^\mu \phi \right\}
 - \e^{2\phi} V\left( \e^{-\phi}, \Theta \right) \right\} \, .
\end{align}
This action (\ref{FRGBg4b}) may have ghost degrees of freedom due to the existence of $\Theta$.
We might eliminate the ghost degrees of freedom by writing $\Theta$ as
$\Theta = \e^\theta$ and add a constraint to the action
(\ref{FRGBg4b}) by using the Lagrange multiplier field $\lambda$, in the following way,
\begin{align}
\label{FRGBg4c}
S =& \int d^4 x \sqrt{-g} \left\{ \frac{1}{2\kappa^2} \left( R
 - \frac{3}{2}\partial^\sigma \phi \partial_\sigma \phi
 - \lambda\left( \partial_\mu \theta \partial^\mu \theta + \mu^2 \right)
\right) \right. \nn
& \left. + \e^\theta \mathcal{G} - \e^\theta\partial_\mu \theta \left\{
4 \left( R^{\mu\nu} - \frac{1}{2} g^{\mu\nu} R \right) \partial_\nu \phi
+ 2 \left( \partial^\mu \phi \Box \phi
 - \left( \nabla_\nu \nabla^\mu \phi\right) \partial^\nu \phi \right)
+ \partial_\nu \phi \partial^\nu \phi \partial^\mu \phi \right\}
 - \e^{2\phi} V\left( \e^{-\phi}, \e^\theta \right) \right\} \, .
\end{align}
Then the scalar fields $\theta$ and $\lambda$ become non-dynamical
degrees of freedom and the dynamical degrees of freedom are
actually the metric and the scalar field $\phi$, as in the
standard $F(R)$ gravity, therefore no-ghost degrees of freedom
occur in the theory.

We should note
\begin{equation}
\label{Gal1}
\epsilon^{\xi\eta\mu\nu} {\epsilon_{\xi\eta}}^{\rho\sigma}
\partial_\mu \theta \partial_\xi \phi \nabla_\nu \nabla_\sigma \phi
= 2 \partial_\theta \left( \partial^\mu \phi \Box \phi
 - \left( \nabla_\nu \nabla^\mu \phi\right) \partial^\nu \phi \right) \, ,
\end{equation}
whose structure appears in the action (\ref{FRGBg4c})
and also the Galileon model \cite{Nicolis:2008in,Deffayet:2009mn,Shirai:2012iw}.
Therefore in the field equations, the term given by the variation of the structure (\ref{Gal1})
does not include the derivative with respect to time $t$ higher than two, which tells that
there could not appear any ghost coming from the higher derivative terms.
As clear from Eq.~(\ref{fG6}), the structure (\ref{Gal1}) appears due to the
combination in the Gauss-Bonnet invariant (\ref{GB}).
Therefore if we consider the action including the combination
$R^2 + b R_{\mu\nu} R^{\mu\nu} + c R_{\mu\nu\rho\sigma} R^{\mu\nu\rho\sigma}$
different from the combination in the Gauss-Bonnet invariant (\ref{GB}) $\left( b \neq - 4c \right)$,
or in more general, the Lagrangian density
$F\left( R, R_{\mu\nu} R^{\mu\nu}, R_{\mu\nu\rho\sigma} R^{\mu\nu\rho\sigma} \right)$,
there appears the ghost.

\section{Field Equations and Formalism in the Ghost-free $F\left( R,\mathcal{G} \right)$ Gravity \label{SecIV}}

Upon varying the action (\ref{FRGBg4c}) with respect to $\lambda$,
$\theta$, $\phi$, and metric $g_{\mu\nu}$, we obtain the following
field equations,
\begin{align}
\label{Eq1}
0=& \partial_\mu \theta \partial^\mu \theta + \mu^2 \, , \\
\label{Eq2}
0=& \frac{1}{\kappa^2} \nabla^\mu \left( \lambda \partial_\mu \theta \right)
+ \e^\theta \mathcal{G} + \e^\theta \left\{
4 \left( R^{\mu\nu} - \frac{1}{2} g^{\mu\nu} R \right) \nabla_\mu \nabla_\nu \phi \right. \nn
& \left. + 2 \left( \left( \Box\phi \right)^2 + \partial^\mu \phi \partial_\mu \Box \phi
 - \left( \nabla_\mu \nabla_\nu \nabla^\mu \phi \right) \partial^\nu \phi
 - \left( \nabla_\nu \nabla^\mu \phi \right) \left( \nabla_\mu \nabla^\nu \phi \right) \right)
+ 2 \left( \nabla_\mu \nabla_\nu \phi \right) \partial^\nu \phi \partial^\mu \phi
+ \partial_\nu \phi \partial^\nu \phi \Box \phi\right\} \, , \\
\label{Eq3}
0=& \frac{3}{2\kappa^2} \Box\phi
+ 4 \left( R^{\mu\nu} - \frac{1}{2} g^{\mu\nu} R \right) \nabla_\nu \nabla_\mu \e^\theta
+ 2 \left( \Box \e^\theta \right) \Box \phi
+ 2 \left( \partial_\mu \e^\theta \right) \partial^\mu \Box \phi
 - 2 \Box \left( \left( \partial_\mu \e^\theta \right) \partial^\mu \phi \right)
+ 2 \nabla^\mu \nabla_\nu \left( \left(\partial_\mu \e^\theta\right)\partial^\nu \phi \right) \nn
& - 2 \nabla^\nu \left( \left( \partial_\mu \e^\theta \right) \nabla_\nu \nabla^\mu \phi \right)
+ 2 \nabla_\nu \left( \left( \partial_\mu \e^\theta \right) \partial^\nu \phi \partial^\mu \phi \right)
+ \nabla^\mu \left( \left( \partial_\mu \e^\theta \right) \partial_\nu \phi \partial^\nu \phi \right) \nn
& - 2 \e^{2\phi} V\left( \e^{-\phi}, \e^\theta \right)
+ \e^\phi \frac{\partial V\left( \e^{-\phi}, \e^\theta \right)}{\partial \e^{-\phi}} \, , \\
\label{Eq4}
0 =& - \frac{1}{2\kappa^2} \left( R_{\mu\nu} - \frac{3}{2} \partial_\mu \phi \partial_\nu \phi
 - \lambda \partial_\mu \theta \partial_\nu \theta \right)
+ \frac{1}{2} g_{\mu\nu} \left\{ \frac{1}{2\kappa^2} \left( R
 - \frac{3}{2}\partial^\sigma \phi \partial_\sigma \phi
 - \lambda\left( \partial_\mu \theta \partial^\mu \theta + \mu^2 \right)
\right) \right. \nn
& \left. - \e^\theta\partial_\rho \theta \left\{
4 \left( R^{\rho\sigma} - \frac{1}{2} g^{\rho\sigma} R \right) \partial_\sigma \phi
+ 2 \left( \partial^\rho \phi \Box \phi
 - \left( \nabla_\sigma \nabla^\rho \phi\right) \partial^\sigma \phi \right)
+ \partial_\sigma \phi \partial^\sigma \phi \partial^\rho \phi \right\}
 - \e^{2\phi} V\left( \e^{-\phi}, \e^\theta \right) \right\} \nn
& + 2 \left( \nabla_\mu \nabla_\nu \e^\theta \right)R
 - 2 g_{\mu\nu} \left( \Box \e^\theta \right)R
 - 4 \left( \nabla^\rho \nabla_\mu \e^\theta \right)R_{\nu\rho}
 - 4 \left( \nabla^\rho \nabla_\nu \e^\theta \right)R_{\mu\rho} \nn
& + 4 \left( \Box \e^\theta \right)R_{\mu\nu}
+ 4g_{\mu\nu} \left( \nabla_{\rho} \nabla_\sigma \e^\theta \right) R^{\rho\sigma}
- 4 \left(\nabla^\rho \nabla^\sigma \e^\theta \right) R_{\mu\rho\nu\sigma} \nn
& - \frac{1}{2} \e^\theta\partial_\mu \theta \left\{
4 R_\nu^{\ \rho} \partial_\rho \phi - 2 R \partial_\nu \phi
+ 2 \left( \partial_\nu \phi \Box \phi
 - \left( \nabla_\rho \nabla_\nu \phi\right) \partial^\rho \phi \right)
+ \partial_\rho \phi \partial^\rho \phi \partial_\nu \phi \right\} \nn
& - \frac{1}{2} \e^\theta\partial_\nu \theta \left\{
4 R_\mu^{\ \rho} \partial_\rho \phi - 2 R \partial_\mu \phi
+ 2 \left( \partial_\mu \phi \Box \phi
 - \left( \nabla_\rho \nabla_\mu \phi\right) \partial^\rho \phi \right)
+ \partial_\rho \phi \partial^\rho \phi \partial_\mu \phi \right\} \nn
& + 2 \left( \partial^\rho \e^\theta \right) R_{\rho\mu} \partial_\nu \phi
+ 2 \left( \partial^\rho \e^\theta \right) R_{\rho\nu} \partial_\mu \phi
 - 2 \left( \partial^\rho \e^\theta \right) \partial_\rho \phi R_{\mu\nu}
 - 2 \nabla_\rho \nabla_\nu \left( \left( \partial^\rho \e^\theta \right) \partial_\mu \phi \right)
 - 2 \nabla_\rho \nabla_\mu \left( \left( \partial^\rho \e^\theta \right) \partial_\nu \phi \right) \nn
& + \Box \left( \left( \partial_\mu \e^\theta \right) \partial_\nu \phi
+ \left( \partial_\nu \e^\theta \right) \partial_\mu \phi \right)
+ g_{\mu\nu} \left( \nabla_\rho \nabla_\sigma + \nabla_\sigma \nabla_\rho \right)
\left( \left( \partial^\rho \e^\theta \right) \partial^\sigma \phi \right)
+ 2 \left( \nabla_\mu \nabla_\nu - g_{\mu\nu} \Box \right) \left( \left(\partial^\rho \e^\theta \right) \partial_\rho \phi \right) \nn
& + \nabla_\mu \left( \left(\partial_\rho \e^\theta \right) \partial^\rho \phi \partial_\nu \phi \right)
+ \nabla_\nu \left( \left(\partial_\rho \e^\theta \right) \partial^\rho \phi \partial_\mu \phi \right)
 - g_{\mu\nu} \nabla_\rho \left( \left(\partial_\sigma \e^\theta \right) \partial^\rho \phi \partial^\sigma \phi \right)
 - \left( \partial_\rho \e^\theta \right) \left(\left(\nabla_\mu \nabla^\rho \phi \right) \partial_\nu \phi
+ \left(\nabla_\nu \nabla^\rho \phi \right) \partial_\mu \phi \right) \nn
& - \nabla^\rho \left(\left( \partial_\mu \e^\theta \right) \partial_\nu \phi \partial_\rho \phi \right)
 - \nabla^\rho \left(\left( \partial_\nu \e^\theta \right) \partial_\mu \phi \partial_\rho \phi \right)
+ \frac{1}{2} \nabla^\rho \left(\left( \partial_\mu \e^\theta \right) \partial_\rho \phi \partial_\nu \phi \right)
+ \frac{1}{2} \nabla^\rho \left(\left( \partial_\nu \e^\theta \right) \partial_\rho \phi \partial_\mu \phi \right) \nn
& + \left( \partial_\rho \e^\theta \right) \partial^\rho \phi \partial_\mu \phi \partial_\nu \phi \, .
\end{align}
We now assume that the background metric is the Friedmann-Robertson-Walker (FRW)
spacetime with a flat spatial part,
\begin{equation}
\label{FRWmetric}
ds^2 = - dt^2 + a(t)^2 \sum_{i=1,2,3} \left( dx^i \right)^2 \, ,
\end{equation}
and we also assume $\theta$ and $\phi$ only depend on the cosmic time $t$.
Then Eq.~(\ref{Eq1}) yields,
\begin{equation}
\label{FRGBa1}
\theta = \pm \mu t \, .
\end{equation}
We absorb the signature $\pm$ into the redefinition of $\mu$ by
assuming that $\mu$ can be negative, hence $\theta = \mu t$.
Then Eqs.~(\ref{Eq2}), and (\ref{Eq3}) take the following forms,
\begin{align}
\label{Eq5}
0=& - \frac{\mu}{\kappa^2} \left( \dot\lambda + 3H \lambda \right)
+ \e^{\mu t} \left[ 24 H^2 \left(\dot H + H^2\right) -12 H^2 \ddot\phi
 -12 \left( 2\dot H + 3 H^2 \right) H \dot \phi + 18 H^2 {\dot\phi}^2 \right. \nn
& \left. + 12 H \ddot\phi \dot\phi +6\dot{H} {\dot\phi}^2
+ 3 \left( \ddot\phi {\dot\phi}^2 + H {\dot\phi}^3 \right) \right] - \e^{2\phi + \mu t}
\left. \frac{\partial V\left( \e^{-\phi}, \e^\theta \right)}{\partial \e^\theta} \right|_{\theta = \mu t} \, , \\
\label{Eq6}
0=& - \frac{3}{2\kappa^2} \left( \ddot{\phi} +3H \dot{\phi} \right)
+\left[ \mu \left( 9 H {\dot{\phi}}^2 + 36 H^2 \dot{\phi} - 36 H^3 +6 \dot{\phi} \ddot{\phi}
+12 \dot{H} \dot{\phi} +12 H \ddot{\phi} -24 H \dot{H} \right) \right. \nn
& \left. +\mu^2\left( 3 {\dot{\phi}}^2 +12H \dot{\phi} -12 H^2 \right)
\right] \e^{\mu t} - 2 \e^{2\phi} V\left( \e^{-\phi}, \e^{\mu t} \right)
+ \left. \e^\phi \frac{\partial V\left( \e^{-\phi}, \e^\theta \right)}{\partial \e^{-\phi}} \right|_{\theta = \mu t} \, .
\end{align}
where $H=\dot{a}/a$ and the $(\mu,\nu)=(t,t)$ and
$(\mu,\nu)=(i,j)$ components in Eq. ~(\ref{Eq4}) yield,
\begin{align}
\label{ttHmu}
0=& \frac{1}{\kappa^2} \left( 3 H^2 -\frac34 {\dot{\phi}}^2 -\lambda \mu^2 \right)
+\left( 3 {\dot{\phi}}^3 -36 H^2 \dot{\phi} +18 H {\dot{\phi}}^2 +24 H^3 \right) \mu \e^{\mu t}
 - \left. \e^{2 \phi} V\left( \e^{-\phi}, \e^\theta \right) \right|_{\theta = \mu t} \, , \\
\label{C1}
0= &\frac{1}{\kappa^2} \left( 2 \dot{H} +\frac{3}{4} {\dot{\phi}}^2 + 3 H^2 \right)
+\left( 8 H^2 + 2 {\dot{\phi}}^2 - 8 H \dot{\phi} \right) \mu^2 \e^{\mu t} \nn
& +\left( -8 \dot{H} \dot{\phi} + 4 \dot{\phi} \ddot{\phi} + 16 H^3 - 8 H \ddot{\phi} - {\dot{\phi}}^3
+ 16 H \dot{H} - 12 H^2 \dot{\phi} \right) \mu \e^{\mu t}
 - \left. \e^{2 \phi} V\left( \e^{-\phi}, \e^\theta \right) \right|_{\theta = \mu t} \, .
\end{align}
We should note that all the equations, (\ref{Eq5}), (\ref{Eq6}),
(\ref{ttHmu}), and (\ref{C1}) do not contain higher order
derivatives and the maximum order of derivatives contained is two,
a fact that indicates that the theory is ghost-free.

By combining Eqs.~(\ref{ttHmu}) and (\ref{C1}), we may delete
$V\left( \e^{-\phi}, \e^\theta \right)$ and we can solve the
obtained equation with respect to $\lambda$,
\begin{align}
\label{C3}
\frac{\mu^2}{\kappa^2} \lambda = & - \frac{1}{\kappa^2} \left( 2 \dot{H} + \frac{3}{2} {\dot{\phi}}^2 \right)
 - \left( 8 H^2 + 2 {\dot{\phi}}^2 - 8 H \dot{\phi} \right) \mu^2 \e^{\mu t} \nn
& - \left( -8 \dot{H} \dot{\phi} + 4 \dot{\phi} \ddot{\phi} - 8 H^3 - 8 H \ddot{\phi} - 18 H {\dot{\phi}}^2 - 4 {\dot{\phi}}^3
+ 16 H \dot{H} + 24 H^2 \dot{\phi} \right) \mu \e^{\mu t} \, .
\end{align}
which determines the $t$ dependence of $\lambda$.
Eq.~(\ref{ttHmu}) also determines the $t$ dependence of
$V\left( \e^{-\phi}, \e^\theta \right)$,
\begin{align}
\label{C4}
\left. \e^{2 \phi} V\left( \e^{-\phi}, \e^\theta \right) \right|_{\theta = \mu t}
= &\frac{1}{\kappa^2} \left( 2 \dot{H} +\frac{3}{4} {\dot{\phi}}^2 +3 H^2 \right)
+\left( 8 H^2 + 2 {\dot{\phi}}^2 - 8 H \dot{\phi} \right) \mu^2 \e^{\mu t} \nn
& +\left( -8 \dot{H} \dot{\phi} + 4 \dot{\phi} \ddot{\phi} + 16 H^3 - 8 H \ddot{\phi} - {\dot{\phi}}^3
+ 16 H \dot{H} - 12 H^2 \dot{\phi} \right) \mu \e^{\mu t} \, .
\end{align}
By eliminating $\lambda$ in Eq. ~(\ref{Eq5}) by using Eq.~(\ref{C3}), we get,
\begin{align}
\label{C5}
0=& \frac{1}{\kappa^2} \left( 2 \ddot H + 6 H \dot H + \frac{9}{2}H {\dot\phi}^2 + 3 \dot\phi \ddot\phi \right) \nn
& + \left( 48 H^2 \dot H + 16 {\dot H}^2 + 16 H\ddot H
+ 36 H^3 \dot\phi -8 \ddot{H} \dot{\phi}
 - 12 H^2 \ddot\phi - 16 \dot H \ddot\phi - 8 H \dddot{\phi}
 - 36 H^2 {\dot\phi}^2 \right. \nn
& \left. - 12 \dot H {\dot\phi}^2 - 12 H \dot{\phi} \ddot\phi
+ 4 {\ddot\phi}^2 + 4 \dot{\phi} \dddot{\phi} - 9 H {\dot\phi}^3 -9 {\dot{\phi}}^2 \ddot\phi
\right) \mu \e^{\mu t} \nn
& + \left( 16 H^3 + 32 H \dot H - 16 \dot H \dot\phi
 - 16 H \ddot{\phi} + 8 \dot\phi \ddot\phi - 12 H {\dot{\phi}}^2 - 4 {\dot{\phi}}^3
\right) \mu^2 \e^{\mu t}
+ \left( 8 H^2 + 2 {\dot{\phi}}^2 - 8 H \dot{\phi} \right) \mu^3 \e^{\mu t} \nn
& - \mu \e^{2\phi + \mu t } \left. \frac{\partial V\left( \e^{-\phi}, \e^\theta \right)}{\partial \e^\theta} \right|_{\theta = \mu t} \, ,
\end{align}
We should note that all the equations (\ref{Eq5}), (\ref{Eq6}),
(\ref{ttHmu}), and (\ref{C1}) are not independent equations.
For example, we consider the derivative of Eq.~(\ref{C1}) with respect to $t$,
\begin{align}
\label{C6}
0= &\frac{1}{\kappa^2} \left( 2 \ddot{H} +\frac{3}{2} \dot{\phi} \ddot\phi + 6 H \dot H \right)
+\left( 16 H \dot H + 4 \dot{\phi} \ddot\phi - 8 \dot H \dot{\phi} - 8 H \ddot\phi \right) \mu^2 \e^{\mu t}
+\left( 8 H^2 + 2 {\dot{\phi}}^2 - 8 H \dot{\phi} \right) \mu^3 \e^{\mu t} \nn
& +\left( -8 \ddot{H} \dot{\phi} - 8 \dot H \ddot\phi + 4 {\ddot\phi}^2 + 4 \dot{\phi} \dddot{\phi} + 48 H^2 \dot H
 - 8 \dot H \ddot\phi - 8 H \dddot{\phi} - 3 {\dot{\phi}}^2 \ddot\phi
+ 16 {\dot H}^2+ 16 H \ddot{H} \right. \nn
& \left. - 24 H\dot H \dot\phi - 12 H^2 \ddot{\phi} \right) \mu \e^{\mu t}
+ \left( -8 \dot{H} \dot{\phi} + 4 \dot{\phi} \ddot{\phi} + 16 H^3 - 8 H \ddot{\phi} - {\dot{\phi}}^3
+ 16 H \dot{H} - 12 H^2 \dot{\phi} \right) \mu^2 \e^{\mu t} \nn
& + 192 H \dot H \left( \dot{H} +H^2 \right) \e^{\mu t}
+ 96 H^2 \left( \ddot{H} +2 H \dot H \right) \e^{\mu t} + 96 \mu H^2 \left( \dot{H} +H^2 \right) \e^{\mu t} \nn
& - \left. 2\dot\phi \e^{2 \phi} V\left( \e^{-\phi}, \e^\theta \right) \right|_{\theta = \mu t}
+ \left. \dot\phi \e^{2 \phi} \frac{\partial V\left( \e^{-\phi}, \e^\theta \right)}{\partial \e^{-\phi}} \right|_{\theta = \mu t}
 - \left. \mu \e^{2 \phi + \mu t} \frac{\partial V\left( \e^{-\phi}, \e^\theta \right)}{\partial \e^\theta} \right|_{\theta = \mu t} \, .
\end{align}
If we eliminate $V\left( \e^{-\phi}, \e^\theta \right)$ in Eq.~(\ref{C6}) by using Eqs.~(\ref{Eq6}) and (\ref{C5}),
we find that the right hand side vanishes identically.
Therefore all the equations (\ref{Eq5}), (\ref{Eq6}), (\ref{ttHmu}), and (\ref{C1})
are not independent equations.

\section{Reconstruction of Several Cosmologies \label{SecV}}

In this section, we shall use the framework we developed in the
previous section as a reconstruction technique in order to realize
several models of cosmological interest.
This reconstruction technique is based simply on fixing the behaviors or time
dependencies of $H$ and $\phi$ as follows,
\begin{equation}
\label{R0}
H=H(t)\, , \quad \phi=\phi(t) \, ,
\end{equation}
and eventually we determine which model can realize the given cosmological evolutions.
As discussed in the last section, all the equations (\ref{Eq5}), (\ref{Eq6}), (\ref{ttHmu}),
and (\ref{C1}) are not independent equations.
Since we solve the equations with
respect to $\lambda$ in (\ref{C3}), we consider Eq.~(\ref{C4}),
which is equivalent to Eq.~(\ref{ttHmu}), and Eq.~(\ref{C5}),
which are independent equations which do not include $\lambda$.
As a working hypothesis, we assume that $V\left( \e^{-\phi}, \e^\theta \right)$
is given by a sum of a $\phi$-dependent part
and of a $\theta$-dependent part as follows,
\begin{equation}
\label{R1}
V\left( \e^{-\phi}, \e^\theta \right) = V_\Phi\left( \e^{-\phi} \right)
+ V_\Theta\left( \e^\theta \right) \, .
\end{equation}
Then Eq.~(\ref{C5}) indicates,
\begin{align}
\label{R2}
V_\Theta\left( \e^\theta \right)=& \int^{\frac{\theta}{\mu}} dt \e^{-2\phi} \left[ \frac{1}{\kappa^2} \left( 2 \ddot H + 6 H \dot H
+ \frac{9}{2}H {\dot\phi}^2 + 3 \dot\phi \ddot\phi \right) \right. \nn
& + \left( 48 H^2 \dot H + 16 {\dot H}^2 + 16 H\ddot H
+ 36 H^3 \dot\phi -8 \ddot{H} \dot{\phi}
 - 12 H^2 \ddot\phi - 16 \dot H \ddot\phi - 8 H \dddot{\phi}
 - 36 H^2 {\dot\phi}^2 \right. \nn
& \left. - 12 \dot H {\dot\phi}^2 - 12 H \dot{\phi} \ddot\phi
+ 4 {\ddot\phi}^2 + 4 \dot{\phi} \dddot{\phi} - 9 H {\dot\phi}^3 -9 {\dot{\phi}}^2 \ddot\phi
\right) \mu \e^{\mu t} \nn
& \left. + \left( 16 H^3 + 32 H \dot H - 16 \dot H \dot\phi
 - 16 H \ddot{\phi} + 8 \dot\phi \ddot\phi - 12 H {\dot{\phi}}^2 - 4 {\dot{\phi}}^3
\right) \mu^2 \e^{\mu t}
+ \left( 8 H^2 + 2 {\dot{\phi}}^2 - 8 H \dot{\phi} \right) \mu^3 \e^{\mu t} \right] \, ,
\end{align}
and Eq.~(\ref{C4}) implies,
\begin{align}
\label{R3}
V_\Phi\left( \e^{-\phi} \right)
= &\e^{-2\phi} \left[ \frac{1}{\kappa^2} \left( 2 \dot{H} +\frac{3}{4} {\dot{\phi}}^2 +3 H^2 \right)
+\left( 8 H^2 + 2 {\dot{\phi}}^2 - 8 H \dot{\phi} \right) \mu^2 \e^{\mu t} \right. \nn
& \left. \left. +\left( -8 \dot{H} \dot{\phi} + 4 \dot{\phi} \ddot{\phi} + 16 H^3 - 8 H \ddot{\phi} - {\dot{\phi}}^3
+ 16 H \dot{H} - 12 H^2 \dot{\phi} \right) \mu \e^{\mu t} \right] \right|_{t=t(\phi)}
 - V_\Theta\left( \e^{\mu t(\phi)} \right)\, .
\end{align}
Here we have assumed $\phi=\phi(t)$ can be solved with respect to $t$ as $t=t(\phi)$.
Then for an arbitrary cosmological evolution
given by $H=H(t)$ and $\phi=\phi(t)$, if we construct the
potential as in (\ref{R1}), (\ref{R2}), and (\ref{R3}), the
evolution of $H$ and $\phi$ can be realized. Note that by making
the inverse conformal transformation we performed below Eq.~(\ref{FRGBg4}),
one may obtain the above potentials in the
original frame, but the resulting theory is too complicated to be
analyzed, so we omit the details.

If we are only interested in the in realizing a specific cosmology
with a given Hubble rate $H$, we may choose $\phi=\phi(t)$ to have
a specific but simple form like,
\begin{equation}
\label{R4}
\phi = \phi_0 t \, ,
\end{equation}
with a constant $\phi_0$. Then Eqs.~(\ref{R2}) and (\ref{R3}) are
simplified and take the following form,
\begin{align}
\label{R5}
V_\Theta\left( \e^\theta \right)=& \int^{\frac{\theta}{\mu}} dt \e^{-2\phi_0 t} \left[ \frac{1}{\kappa^2} \left( 2 \ddot H + 6 H \dot H
+ \frac{9}{2}H {\phi_0}^2 \right) \right. \nn
& + \left( 48 H^2 \dot H + 16 {\dot H}^2 + 16 H\ddot H + 36 H^3 \phi_0 -8 \ddot{H} \phi_0 - 36 H^2 {\phi_0}^2
 - 12 \dot H {\phi_0}^2 - 9 H {\phi_0}^3 \right) \mu \e^{\mu t} \nn
& \left. + \left( 16 H^3 + 32 H \dot H - 16 \dot H \phi_0 - 12 H {\phi_0}^2 - 4 {\phi_0}^3 \right) \mu^2 \e^{\mu t}
+ \left( 8 H^2 + 2 {\phi_0}^2 - 8 H \phi_0 \right) \mu^3 \e^{\mu t} \right] \, , \\
\label{R6}
V_\Phi\left( \e^{-\phi} \right)
= &\e^{-2\phi_0 t}
\left[ \frac{1}{\kappa^2} \left( 2 \dot{H} +\frac{3}{4} {\phi_0}^2 +3 H^2 \right)
+\left( 8 H^2 + 2 {\phi_0}^2 - 8 H \phi_0 \right) \mu^2 \e^{\mu t} \right. \nn
& \left. \left. +\left( -8 \dot{H} \phi_0 + 16 H^3 - {\phi_0}^3
+ 16 H \dot{H} - 12 H^2 \phi_0 \right) \mu \e^{\mu t} \right] \right|_{t=\frac{\phi}{\phi_0}}
 - V_\Theta\left( \e^{\frac{\mu \phi}{\phi_0}} \right)\, .
\end{align}
Furthermore by choosing,
\begin{equation}
\label{R7}
2\phi_0=\mu\, .
\end{equation}
we obtain,
\begin{align}
\label{R8}
V_\Theta\left( \e^\theta \right)
=& \int^{\frac{\theta}{\mu}} dt \left[ \frac{\e^{-\mu t}}{\kappa^2} \left( - 2 \mu \dot H - 3 \mu H^2
+ \frac{9}{8}H \mu^2 \right)
+ 34 \mu^2 H^3 - \mu^3 H^2 - \frac{65}{8} \mu^4 H \right] \nn
& + \left. \frac{\e^{-\mu t}}{\kappa^2} \left( 2 \dot H + 3 H^2 \right)
+ \left( 16 H^3 + 16 H\dot H \right) \mu
+ \left( - 4 \dot{H} + 16 H^2 \right) \mu^2 - 11 H \mu^3 \right|_{t=\frac{\theta}{\mu}} \, , \\
\label{R9}
V_\Phi\left( \e^{-\phi} \right)
= & \left. \left[ \frac{3 \mu^2 \e^{- \mu t}}{16 \kappa^2}
 - 14 \mu^2 H^2 + 7 H \mu^3
 - \frac{3\mu^4}{8} \right] \right|_{t=\frac{2\phi}{\mu}} \nn
& - \int^{\frac{2\phi}{\mu}} dt \left[ \frac{\e^{-\mu t}}{\kappa^2} \left( - 2 \mu \dot H - 3 \mu H^2
+ \frac{9}{8}H \mu^2 \right)
+ 34 \mu^2 H^3 - \mu^3 H^2 - \frac{65}{8} \mu^4 H \right] \, .
\end{align}
By using Eqs.~(\ref{R8}) and (\ref{R9}), we may consider the
realization of several cosmological evolution.
As a first example, we may consider,
\begin{equation}
\label{R10}
H=\frac{H_0 \e^{-\mu t}}{1 + \e^{-\mu t}} = H_0 \left( 1 - \frac{1}{1 + \e^{-\mu t}} \right)\, .
\end{equation}
When $t\to - \infty$, $H$ goes to a constant $H\to H_0$, which may
describe the de Sitter inflationary evolution.
On the other hand, we find $H\to 0$ when $t\to + \infty$.
Therefore, inflation ends when
$t\sim 0$. We now ignore the contribution from the perfect matter
fluids that may be present, for the behavior of $H$ given in (\ref{R10}).
Then if the matter fluids are coupled with the
scalar fields $\theta$ and/or $\phi$, the matter fluids may
produce non-trivial effects at $t\sim 0$ and the behavior of $H$
could be changed so that the matter or radiation dominated phase
could be generated. Then Eqs.~(\ref{R8}) and (\ref{R9}) yield,
\begin{align}
\label{R15}
V_\Theta\left( \e^\theta \right)
= & \left( \frac{6 H_0^2}{\kappa^2} - \frac{9 \mu H_0}{8\kappa^2} \right) \e^{-\theta}
+ \left( \frac{2H_0 \mu}{\kappa^2} - \frac{6 H_0^2}{\kappa^2} + \frac{9 \mu H_0}{8\kappa^2} - 34 \mu H_0^3
+ \mu^2 H_0^2 + \frac{65}{8} \mu^3 H_0 \right) \ln \left( 1 + \e^{-\theta} \right) \nn
& + \left( \frac{2H_0 \mu}{\kappa^2} - \frac{6 H_0^2}{\kappa^2} + \frac{9 \mu H_0}{8\kappa^2} - 102 \mu H_0^3
 - 2 \mu^2 H_0^2 - \frac{65}{8} \mu^3 H_0 \right) \ln \left( 1 + \e^{-\theta} \right) \nn
& + \frac{2 H_0 \mu}{\kappa^2} - \frac{6 H_0^2}{\kappa^2} + 16 H_0^3 \mu + 16 H_0^2 \mu^2 - 11 H_0 \mu^3 \nn
& + \left( - \frac{2H_0 \mu}{\kappa^2} + \frac{6 H_0^2}{\kappa^2} - 116 \mu H_0^3 + 33 \mu^2 H_0^2 + 7 H_0 \mu^3 \right) \frac{1}{1 + \e^{-\theta}} \nn
& + \left( \frac{2 H_0 \mu}{\kappa^2} - \frac{3 H_0^2}{\kappa^2} + 65 H_0^3 \mu - 16 H_0^2 \mu^2 + 4 H_0 \mu^3 \right) \frac{1}{\left( 1 + \e^{-\mu t} \right)^2}
+ \left( 16 H_0^3 \mu + 16 H_0^2 \mu^2 \right) \frac{1}{\left(1 + \e^{-\theta}\right)^3} \, , \\
\label{16}
V_\Phi\left( \e^{-\phi} \right)
=& \left( - \frac{3 H_0^2}{\kappa^2} + \frac{9 \mu H_0}{8\kappa^2} + \frac{3 \mu^2}{16 \kappa^2} \right) \e^{-2\phi} \nn
& - \left( \frac{2H_0 \mu}{\kappa^2} - \frac{6 H_0^2}{\kappa^2} + \frac{9 \mu H_0}{8\kappa^2} - 34 \mu H_0^3
+ \mu^2 H_0^2 + \frac{65}{8} \mu^3 H_0 \right) \ln \left( 1 + \e^{-2\phi} \right) \nn
& - \left( \frac{2H_0 \mu}{\kappa^2} - \frac{3 H_0^2}{\kappa^2} - 68\mu H_0^3 - 27 \mu^2 H_0^2 - 7 \mu^3 H_0 \right) \frac{1}{1 + \e^{-2\phi}}
 - \frac{17 \mu H_0^3 - 14 \mu^2 H_0^2} {\left(1 + \e^{-2\phi}\right)^2} \, .
\end{align}
Hence the above functional forms for the model at hand, realize
the cosmological evolution of Eq.~(\ref{R10}).

We may consider a more complicated cosmological evolution than
that in Eq.~(\ref{R10}). Since observationally it seems that there
is a tension between the value of the Hubble constant inferred
from small redshifts, as in the observation of Type Ia supernova
(SNIa) calibrated by Cepheid observations $H=H_\mathrm{l}\sim 73
\mathrm{km}\,\mathrm{s}^{-1}\,\mathrm{Mpc}^{-1}$
\cite{Riess:2020fzl} and that from large redshifts, as the cosmic
microwave background (CMB) $H=H_\mathrm{e}\sim 67
\mathrm{km}\,\mathrm{s}^{-1}\,\mathrm{Mpc}^{-1}$
\cite{Planck:2018vyg}. Although the tension might come from the
uncertainties of the Cepheid calibration (see
\cite{Mortsell:2021nzg,Perivolaropoulos:2021bds} for example), a
solution to solve the tension is to introduce an early dark energy
occurring after the inflationary era
\cite{Oikonomou:2020qah,Nojiri:2019fft}. In our model, instead of
(\ref{R10}), we may consider,
\begin{equation}
\label{R16}
H=\frac{H_0 \e^{-\mu t}}{1 + \e^{-\mu t}} + \frac{H_\mathrm{e} - H_\mathrm{l} \e^{-\mu \left(t - t_0\right)}}{1 + \e^{-\mu \left(t - t_0\right)}}
+ H_\mathrm{l} \, .
\end{equation}
Here $t_0$ is a positive constant and $H_0$ is mach larger than $H_\mathrm{e}$ and $H_\mathrm{l}$.
We also assume $t_0 \gg \frac{1}{\mu}$. Then during the era when $1/\mu \ll t \ll t_0$,
$H$ behaves as a constant $H\sim H_\mathrm{e} - H_\mathrm{l} + H_\mathrm{l} = H_\mathrm{e}$, which could correspond to
the Hubble rate at the clear up of the Universe, when the CMB was generated.
In this epoch, our model play the role of the early dark energy.
On the other hand, when $t \gg t_0 \left(\gg 1/\mu\right)$,  we find $H\sim H_\mathrm{l}$, which correspond to the present
Hubble rate.
Therefore the model (\ref{R16}) can describe all of the inflation, the early dark energy, and the dark energy in the present Universe.

As another example, we may consider a model mimicking the
$\Lambda$CDM model,
\begin{equation}
\label{R17}
H=\frac{1}{l} \coth \left( \frac{3t}{2l} \right) \, .
\end{equation}
Here $l$ is the length of the effective de Sitter radius.
Specifically if we choose,
\begin{equation}
\label{R18}
\mu = \frac{3}{l}\, ,
\end{equation}
we can perform the integrations in (\ref{R8}) and (\ref{R9}) and
we obtain,
\begin{align}
\label{R23_}
V_\Theta\left( \e^\theta \right)
=& \frac{\mu^2}{\kappa^2} \left( -\frac{3}{4} \theta +\frac{25}{24 \e^{\theta}} +\frac{3}{4} \ln \left( \e^{\theta}-1 \right) \right) \nn
& + \mu^4 \left( \frac{289}{216} \theta -\frac{313}{108} \ln \left( \e^{\theta}-1 \right)
+\frac{-35 \, \e^{3\theta} +47 \, \e^{2\theta} +43\, \e^{\theta} -119 }{27 \left( \e^{\theta}-1 \right)^3} \right) \, , \\
\label{R24_}
V_\Phi\left( \e^{-\phi} \right)
= & \frac{\mu^2}{\kappa^2} \left( - \frac{483}{16} \e^{-2\phi} - \frac{411}{4} \ln \left( 1 - \e^{-2\phi} \right) - \frac{96}{1 - \e^{-2\phi}} \right) \nn
& + \mu^4 \left( \frac{313}{108} \ln \left( \e^{2 \phi} -1 \right) -\frac{289}{108} \phi
+\frac{87 \e^{4\phi} +482 \e^{2\phi} -825 }{216 \left( \e^{2\phi}-1 \right)^2}
\right) \, .
\end{align}
Therefore the model can mimic the $\Lambda$CDM model without the
presence of a cold dark matter perfect fluid.

We can restore the form of the function $F\left( R, \mathcal{G} \right)$ for this model in the limit $t \rightarrow \infty$.
The equations \eqref{FRGBg3} can be rewritten as
\begin{align}
\label{RR}
\frac{R}{2\kappa^2} =& -\frac{\mu^2}{\kappa^2} \frac{\Phi \left( 25\Phi^2 +11 \right)}{24 \left( \Phi^2 -1 \right)}
+\mu^4\left( \frac{289 \Phi^6 +927 \Phi^4 -1041 \Phi^2 +337}{108 \Phi \left( \Phi^2 -1 \right)^3 } \right) \, ,
\quad 1 > \Phi >0 \, , \\
\label{GG}
\mathcal{G} =&-\frac{\mu^2}{\kappa^2 \Theta^2} \frac{ \left( 7\Theta -25 \right)}{24 \left( \Theta -1 \right)}
 -\frac{\mu^4 \left( 337 \Theta^4 -1186 \Theta^3 +1584 \Theta^2 -1982 \Theta -289 \right)}{216 \Theta \left( \Theta -1 \right)^4} \, , \quad \Theta >1 \, .
\end{align}
In the limit $\Phi \rightarrow 0, \ \Theta \rightarrow \infty \ (\mbox{this corresponds to} \ \mu >0, \ t \rightarrow \infty)$, we get
\begin{align}
\label{RRR}
\Phi =& -\frac{337 \mu^4 \kappa^2}{54} \frac{1}{R}
+ \frac{113569 \mu^{10} \kappa^4 \left( 20 \kappa^2 \mu^2 +33 \right)}{104976} \frac{1}{R^3} + \mathcal{O}\left( \frac{1}{R^5} \right)\, , \quad
\left( R \rightarrow -\infty \right) \, , \\
\label{GGG}
\Theta =& -\frac{337 \mu^4 \kappa^2}{216} \frac{1}{\mathcal{G}}
+\frac{9 \left( 18\mu^2 \kappa^2 +7 \right)}{337 \mu^2 \kappa^2} + \mathcal{O} \left( \mathcal{G} \right)\, , \quad
\left( \mathcal{G} \rightarrow 0 \right) \, .
\end{align}
Substituting these expressions into equation \eqref{FRGBg4}, we obtain in the limit $\Phi \rightarrow 0, \ \Theta \rightarrow \infty$
\begin{align}
\label{F22}
F\left( R, \mathcal{G} \right) =& -\frac{337 \mu^4}{432} \ln\left(\frac{R^4}{(2 \kappa^2)^4} \mathcal{G}^2\right)
+\frac{\mu^4}{216} \left( 4044\ln(\mu) -2359\ln(2) -3033\ln(3) +1011\ln(337)+818 \right)
\nonumber \\ &
 -\frac{113569}{419904}\mu^{10}\kappa^2 \left(20 \mu^2 \kappa^2 +33\right)\frac{1}{R^2}
+\left( \frac{162}{337} +\frac{63}{337 \mu^2 \kappa^2}\right)\mathcal{G}
+ \mathcal{O}\left( \frac{1}{R^4} \right) + \mathcal{O} \left( \mathcal{G}^2 \right) \, .
\end{align}

\section{Conclusions \label{SecVI}}

In this work we addressed the ghost issue of $F\left( R,\mathcal{G} \right)$
gravity, which is plagued with ghost degrees of freedom.
These ghost degrees of freedom make the presence at any level that the
theory is considered, and may arise even at the level of
cosmological perturbations, where superluminal modes may occur.
If we consider the quantum theory, the ghosts generate the negative norm states,
which give the negative probabilities, and therefore the ghosts are physically inconsistent.
The ghosts are due to the presence of higher derivatives
eventually in the theory, higher than two at the equations of motion level.
Thus, due to the fact that these theories often
provide viable descriptions of inflation and dark energy, we
provided a theoretical framework in which the ghost degrees of
freedom disappear.
Specifically we introduced two auxiliary scalar
fields, and by using the Lagrange multiplier technique, we
generated a ghost free $F\left( R,\mathcal{G} \right)$ theory in the Einstein
frame, with the only dynamical field being one scalar field.
Accordingly, we derived the field equations of the ghost free
$F\left( R,\mathcal{G} \right)$ gravity for a general metric, and for the flat FRW metric.
The field equations can be viewed as a general
reconstruction technique, in which one may specify the Hubble
rate, the scalar field evolution or even both the previous two,
and by using the field equations one may discover which model can
generate such an evolution.
We used the reconstruction technique
to realize several cosmological evolutions of interest, such as a
de Sitter, an early dark energy evolution and an evolution
mimicking the $\Lambda$CDM evolution.
In a future work, we shall
develop further the formalism in order to study in a quantitative
way the inflationary era and the dark energy era, in this work we
aimed to present the essential features of a ghost-free
$F\left( R,\mathcal{G} \right)$ gravity in the Einstein frame.

\section*{Acknowledgments}

This work was supported by MINECO (Spain), project
PID2019-104397GB-I00 (SDO). This work is also supported by the
JSPS Grant-in-Aid for Scientific Research (C) No. 18K03615 (S.N.).
The work of A.P. was partly funded by the Russian Foundation for
Basic Research grant No. 19-02-00496. The work of A.P was also
funded by the development program of the Regional Scientific and
Educational Mathematical Center of the Volga Federal District,
agreement N 075-02-2020. This paper has been supported by the
Kazan Federal University Strategic Academic Leadership Program.

\end{document}